\documentclass[showpacs,onecolumn,superscriptaddress]{revtex4}
\usepackage{amssymb}
\usepackage{amsmath}
\usepackage{bm}
\usepackage{graphics}
\usepackage{natbib}

\newcommand{\sss}[1]{{\scriptscriptstyle #1}}
\begin{document}
\title{Dirac equation description on the electronic states and
magnetic properties of a square graphene quantum dot}
\author{Changlin Tang}
\affiliation{Department of Physics,  Jilin University, Changchun
130023, PR China}
 \affiliation{State Key Lab of Structural Chemistry, Fujian Institute of Research
on the Structure of Matter and Graduate School of Chinese Academy of
Sciences, Fuzhou 350002, PR China}
\author{Weihua Yan}
\affiliation{Department of Physics,  Jilin University, Changchun
130023, PR China}
\author{Yisong Zheng}
\email[Corresponding author. Email address: ]{zys@jlu.edu.cn}
\affiliation{Department of Physics,  Jilin University, Changchun
130023, PR China}
\author{Guangshe Li}
\affiliation{State Key Lab of Structural Chemistry, Fujian Institute
of Research on the Structure of Matter and Graduate School of
Chinese Academy of Sciences, Fuzhou 350002, PR China}
\author{Liping Li}
\affiliation{State Key Lab of Structural Chemistry, Fujian Institute
of Research on the Structure of Matter and Graduate School of
Chinese Academy of Sciences, Fuzhou 350002, PR China}
\date{\today}
\begin{abstract}
Electronic eigen-states of a square graphene quantum dot(GQD)
terminated by both zigzag and armchair edges are derived in the
theoretical framework of Dirac equation. We find that the Dirac
equation can determine the eigen-energy spectrum of a GQD with high
accuracy even if its size is reduced to a few nanometers. More
importantly, from the Dirac equation description we can readily work
out the number and energy gap of the conjugate surface states, which
are intimately associated with the magnetic properties of the GQD.
By using the Hartree-Fock mean field approach, we study the size
dependence of the magnetic ordering formation in this square GQD. We
find that there exists a critical size of the width between the two
zigzag edges to indicate the onset of the stable magnetic ordering.
On the other hand, when such a width increases further, the magnetic
ground state energy of a charge neutral GQD tends to a saturated
value. These results coincide with the previous results obtained
from the first principle calculation. Then, based on the Dirac
equation solution about the surface state, we establish a simple
two-state model which can quantitatively explain the size dependence
of the magnetic ordering in the square GQD.

\end{abstract}
 \maketitle

\bigskip
\section{Introduction}
Graphene has become a subject of intense interest since the
experimental success in fabricating such an atomically thin layer of
graphite\cite{refNovoselov}. The valence electron dynamics in such a
truly two-dimensional material is governed by a massless Dirac
equation. As a result, graphene exhibits many unique electronic
properties\cite{Meyer,refzheng}, in comparison with the conventional
semiconductor materials. From an application point of view, graphene
possesses very high mobility even at room
temperature\cite{refNovoselov2}. Moreover, the planar geometry of
graphene is of advantage to tailor various nanostructures by the
current experimental means, such as the lithographic
techniques\cite{Zhang}. So far, the obtainable graphene
nanostructures include the one-dimensional(1D) nanoribbons\cite{Han}
and zero-dimensional quantum dots\cite{Geim}. These nanostructures
are viewed as the elemental blocks to construct the graphene-based
nano-devices.
\par
Accompanying the extensive investigations on the electronic
properties in bulk graphene. Graphene nanostructures also draw much
attention of theoretical study\cite{refCastro}. First of all, some
theoretical approaches to produce the effective electron confinement
in graphene were proposed\cite{nilsson,trauzette,efetov,martino}
which is a nontrivial problem due to the Klein tunneling of the
carrier in graphene\cite{cheianov,katsnelson}. For example, GQD
structures can be formed by patterning gates on a semiconducting
graphene nanoribbon\cite{trauzette,efetov}, or by using
inhomogeneous magnetic fields\cite{martino}. Then, some device
applications of the graphene nanostructures were suggested, such as
the spin qubits based on the coupled GQDs\cite{trauzette}. In
addition, some electronic properties of graphene nanostructures are
expected to be different from bulk graphene because of the quantum
confinement and the edge effect. For instance, the spontaneous
magnetization is anticipated to emerge in some graphene
nanostructures\cite{pisani,fujita}, which is attributed to the spin
polarized electron occupancy at the zigzag-type edges of the
nanostructures. And such a magnetic ordering has been experimentally
demonstrated\cite{Enoki,Shibayama}. Quite recently, the possible
magnetism of graphene nanostructures with different shapes is
theoretically studied in some
details\cite{Ezawa,Jianga,Fern,Oded,Young}. For example, an
infinitely long graphene nanoribbon with zigzag edges is possible to
behave as a half-metallic material, in which a spin polarized
current can be formed\cite{Young}. Such a property is controlled by
an external electric field, which can tune the asymmetry of the band
structures of the opposite spin electrons. Apart from the 1D
graphene nanoribbon, GQDs with different shapes also exhibit
magnetic orderings, such as square quantum dot and triangular and
hexagonal quantum dots terminated by zigzag
edges\cite{Ezawa,Jianga,Fern,Oded}. Theoretical calculations
indicate that in these zero-dimensional graphene structures the
magnetic ordering is so robust that can be detected at room
temperature. In addition, for the square and
hexagonal\cite{Jianga,Fern} quantum dots, there is a critical size
which marks the onset of the spin-polarized ground state. On the
contrary, for a quantum dot with a size smaller than the critical
size, the ground state is a paramagnetic state.
\par
The aforementioned results about the magnetic properties of graphene
nanostructures are obtained by means of the first-principle
calculation\cite{Jianga} as well as the mean field approximation on
a Hubbard model of a hexagonal lattice. Although these theoretical
approaches can give the reliable results about the electronic
properties of graphene nanostructures, their applicability is
restricted within those structures with relatively small size.
Furthermore, in most cases these approaches can not provide us with
a clear physical picture to explain the numerical results. For
example, an unambiguous explanation about the critical size for the
onset of the magnetic ordering in GQDs is yet lacking. On the other
hand, the Dirac equation description can just compensate for the
disadvantage of the two theoretical approaches mentioned above. As a
theoretical model based on the effective mass approximation, a
massless Dirac equation can well describe the electron properties of
the bulk graphene as well as the graphene nanostructures with
relatively large sizes\cite{Brey,Brey2}. Usually, such a model can
afford analytical results, which are very helpful to explain
intuitively the electronic properties associated with the
relativistic quantum mechanical feature of graphene.
\par
So far, the Dirac equation succeeds in describing the band
structures of graphene nanoribbons with distinct edge
types\cite{Brey,Brey2}. In the present work, we will employ this
model to study the electron states in a square GQD. With an
appropriate boundary condition, we can derive an analytical solution
of the electron eigen-state in such a GQD. Moreover, by a numerical
calculation carried out from the mean field approximation of the
Hubbard model, we investigate the size dependence of the magnetic
property of the GQD. We find that there are not only the critical
size, but also another characteristic size to indicate the
saturation of the magnetization. Namely, it is a relatively larger
size than the critical size, beyond which the spin polarization
energy no longer varies with the further increase of the size of the
GQD. Then, based on the analytical result obtained from the Dirac
equation, we establish a simple theoretical model, which can not
only reveal the physical nature of the emergence of the critical and
saturated sizes, but also provide an simple way to rapidly create
the quantitative result about the critical and saturated sizes, well
agreeing with the numerical result of the mean field approximation.
\par
The rest of this paper is organized as follows: Starting from the
Dirac equation and using appropriate boundary conditions, the
wavefunction and the dispersion relation of the electron eigenstate
of the square GQD are derived in section \ref{theory}. Then the
eigen-energy spectra calculated from the Dirac equation and the
tight binding model are compared. In section \ref{magnet}, the
magnetic property of the GQD is investigated by means of
Hartree-Fock mean filed theory. By establishing a two-state model,
the size dependence of the magnetic ordering is quantitatively
explained. Finally, the main conclusion is briefly summarized in
section \ref{summary}.

\begin{figure}
\begin{center}
\scalebox{0.5}{\includegraphics{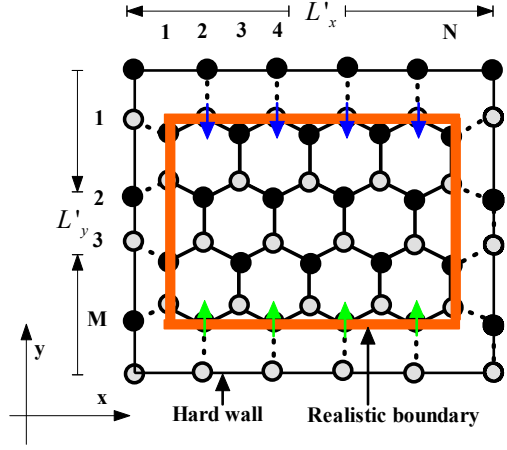}} \caption{A schematic of
the honeycomb lattice of a square GQD terminated by both armchair
and zigzag edges. There are two type of atoms($\bullet$ :A and
$\circ$ : B) The realistic boundary of the GQD is denoted by the
thick lines. A hard wall labeled by the peripheral rectangular
framework consists of the carbon atoms just disconnected from the
atoms at the edges of the GQD. The electron probability amplitudes
at the atoms on the Hard wall must vanish. The size of GQD in x
direction is denoted by $N$ or $L'_x=(N+1)a_0/2$, and in y direction
by $M$ or $L'_y\sqrt{3}a_0(M/2+1/3)$. $a_0$ is the lattice constant.
The upward and downward arrows at the opposite zigzag edges denote
the net spin moments, which indicates the anti-ferromagnetic state.
On the contrary, if the net spin moments at the two zigzag edges
point at the same direction, it indicates the ferromagnetic
state.}\label{1}
\end{center}
\end{figure}

\section{The electronic states of GQD\label{theory}}
The honeycomb lattice of the square quantum dot made of graphene
monolayer is schematically shown in Fig.\ref{1}. The edges of the
square GQD are of two kinds, zigzag edges at the top and bottom, and
armchair ones at the left and right sides.  We assume that the
dangling $\sigma$ bonds at the edges are passivated by hydrogen
atoms. Thus, the behavior of the $\pi$ band electron near Fermi
level is not nontrivially affected by the truncation of the $\sigma$
bond at the GQD edges. Within the effective mass approximation, the
envelop function of the $\pi$ band electron in graphene monolayer
obeys the following Dirac-like equation\cite{refAndo}
\begin{eqnarray}
H\psi=\gamma \begin{bmatrix}0 & -\bm{\hat{k}}_- & 0& 0\\
 -\bm{\hat{k}}_+& 0& 0 & 0\\
 0  & 0 & 0 &  \bm{\hat{k}}_+\\
 0  & 0 & \bm{\hat{k}}_-  & 0
\end{bmatrix}
\begin{bmatrix}\phi_A\\
 \phi_B\\
- \phi_{A}^{'} \\
- \phi_{B}^{'}
\end{bmatrix}
=E
\begin{bmatrix}\phi_A\\
 \phi_B\\
- \phi_{A}^{'} \\
- \phi_{B}^{'}
\end{bmatrix}, \label{dirac}
\end{eqnarray}
where $\gamma$=$\sqrt{3}ta_0/2$ with $t$ being the nearest neighbor
hopping energy. In what follows we use units such that
$\gamma=\hbar=1$. $\bm{\hat{k}}_{\pm}$=$\hat{k}_x\pm i\hat{k}_y$ and
$\hat{k}_{x(y)}=-i\partial_{x(y)}$ is an operator to measure the
momentum deviation from $\bm{K}=(-4\pi/3a_0,0)$ or
$\bm{K}'=(4\pi/3a_0,0)$ point. The four components of the spinor
wavefunction in Eq.(\ref{dirac}) are associated with the total
wavefunction $\Psi(\bm{r})$ by the following relationship.
\begin{equation}
\psi_\mu(\bm{R}_\mu)=e^{i\bm{K}\bm{R}_\mu}\phi_\mu(\bm{R}_\mu)+e^{i\bm{K}'\bm{R}_\mu}\phi'_\mu(\bm{R}_\mu),
\hspace{5mm} \mu=A, B; \label{total wavefunction}
\end{equation}
and
\begin{equation}
\Psi(\bm{r})=\sum_{\mu=A,B}\sum_{\bm{R}_\mu}
\psi_\mu(\bm{R}_\mu)\xi(\bm{r}-\bm{R}_\mu)
\end{equation}
 where $\xi(\bm{r}-\bm{R}_\mu)$ is the carbon atomic wavefunction centered at
 $\bm{R}_\mu$. And $\psi_\mu(\bm{R}_\mu)$ denotes the probability amplitude of the valence electron
 appearing in the vicinity of this carbon atom.
 For the bulk graphene, by solving Eq.(\ref{dirac}) we can obtain
 the electron eigenstate which has the linear dispersion relation
$\varepsilon=E/\gamma=sk$, and s=$\pm1$ denoting the conduction and
valence bands, respectively.
\par
As for the present square GQD structure, the $\pi$ band electron
obeys the same Dirac equation, but is subject to the following
boundary conditions. At the zigzag edges
\begin{eqnarray}
 \phi_B (y=0)=\phi_B' (y=0)=\phi_A
(y=L'_y)=\phi_A' (y=L'_y)=0, \label{zigzag boundary}
\end{eqnarray}
and at the armchair edges
\begin{equation}
\phi_\mu (x=0)=\phi_\mu' (x=0), \;\; \phi_\mu (x=L'_x )=e^{i8\pi
L'_x/3}\phi_\mu' (x=L'_x ).  \label{armchair boundary}
\end{equation}
These boundary conditions have been successfully used to work out
the band structures of the graphene nanoribbons with different
edges\cite{Brey} and the energy spectrum of
GQDs\cite{trauzette,efetov}. They originate from the requirement
that the electron probability amplitude at the hard walls around GQD
must vanish. Combining Eq.(\ref{dirac}) with Eqs.(\ref{zigzag
boundary}-\ref{armchair boundary}), we can derive the eigen solution
of electron states in the square GQD. It is given by
\begin{eqnarray}
\Phi=\begin{bmatrix}\phi_A\\
 \phi_B\\
- \phi_{A}^{'} \\
- \phi_{B}^{'}
\end{bmatrix}
=C\begin{bmatrix}\frac{1}{\varepsilon_n}(-p_n\sin({qy})e^{ip_nx}+q\cos({qy})e^{ip_nx})\\
 \sin({qy})e^{ip_nx}\\
\frac{1}{\varepsilon_n}(p_n\sin({qy})e^{-ip_nx}-q\cos({qy})e^{-ip_nx}) \\
-\sin({qy})e^{-ip_nx}
\end{bmatrix} \label{wavefunction}
\end{eqnarray}
 The corresponding eigen-energy is given by
\begin{eqnarray}
 \varepsilon_n=s\sqrt{p_n^2+q^2}. \label{energy2}
 \end{eqnarray}
Although it takes the same form as the dispersion relation of the
bulk graphene, in the present square GQD the wavevectors $p_n$ and q
in x and y directions are both discrete. $p_n$ is given by
\begin{eqnarray}
p_n=\frac{2n\pi}{(N+1)a_0}-\frac{2\pi}{3a_0}, \; \;
n=0,\pm1,\pm2\cdots \label{wavevector}
\end{eqnarray}
Corresponding to a given $p_n$, $q$ is determined by a
transcendental equation
\begin{eqnarray}
q=p_n\tan{qL'_y}. \label{transcendental equation}
\end{eqnarray}
\par
By analyzing the above equation we find that the electron state with
an imaginary wavevector $q=i|q|$ is allowed in the region
$p_n>1/L'_y$, In such a state the electron wavefunction is localized
in the vicinity of the zigzag edges. Accordingly, it is named as a
surface state. In contrast to the surface state, we call a state
with a real $q$ as the confined state. The normalization coefficient
of the spinor wavefunction for the confined state is
\begin{equation}
C=(2L'_yL'_x-\sin{(2p_nL'_x)}L'_y/p_n)^{-1/2},
\end{equation}
whereas for the surface state it is given by
\begin{equation}
C=((2L'_x-\sin{(2p_nL'_x)}/p_n)(\sinh{(2qL'_y)}/q-2L'_y)/2)^{-1/2}.
\end{equation}
The eigenstate of the GQD shown in Eq.(\ref{wavefunction}) can be
understood in the following way. If the wavefunction is rewritten in
a form $\Phi=\Phi^{p_n}_K+\Phi^{-p_n}_{K'}$ with
$\Phi^{p_n}_K=[\phi_A, \phi_B, 0, 0]^T$ and
$\Phi^{-p_n}_{K'}=[0,0,\phi'_A,
 \phi'_B]^T$, we can immediately find that $\Phi^{p_n}_{K}$ and
 $\Phi^{-p_n}_{K'}$ are just the wavefunctions of the eigenstates of
 an infinitely long zigzag nanoribbon\cite{Brey} with the free wavevectors $p_n$ and $-p_n$
 respectively. Therefore, the eigenstate of the GQD consists of
 the linear combination of the eigenstates of the zigzag nanoribbon in K and K' valleys.
 The boundary condition in the armchair direction admixes K and K' valleys.
 According to such an argument,
 we should restrict the possible surface states in the wavevector range
  $1/L'_y< p_n \leq
 \pi/3a_0$, where $\pi/3a_0$ is just the midpoint between K and K'
valleys in $x$ direction. When $p_n$ is beyond such a value, there
is no longer any new surface state due to the valley admixing.
Considering such a wavevector limit, we can readily determine the
number of the surface states by simply counting the number of the
discrete wavevectors $p_n$ in this range. For example, for a GQD of
size $N=13$ and $M=10$(denoted for short as N13M10), the allowed
$p_n$ are $\pi/21a_0$ and $4\pi/21a_0$. Therefore, the total number
of the surface states in valence and conduction bands are $2N_t=4$,
where $N_t$ is the number of the allowed $p_n$ in the range of
$1/L'_y< p_n \leq \pi/3a_0$.  Moreover, from this method we can
infer that when $N<7$ no surface state survives, independent of the
value of $M$. We will see below that the surface states are
responsible for the magnetic property of the GQD. Such a simple way
to determine the number and energy of the surface states is helpful
for us to explain intuitively the numerical result about the
magnetic property of the GQD. In addition, it should be noted that
in the following discussion, we only consider the GQD with odd $N$
and even $M$. Such a case corresponds to a GQD without any carbon
atom at the boundary connected to the GQD by a single $\pi$ bond.
Finally, although the Dirac equation can give an analytical
description of the electron eigenstate in the square GQD, we have to
point out that its applicability should be strictly restricted
within the linear dispersion region of the $\pi$ band of graphene.
It is known that the energy scope of the linear dispersion region in
the $\pi$ band is about $t/3$, away from the Dirac point.
Accordingly, in the two dimensional k-space the linear dispersion
region forms a circle around the K or K' point with a radius equal
to $2/(3\sqrt{3}a)$. Therefore, if the energy of an eigenstate of
the square GQD is much lower than $t/3$, it can be well described by
the Dirac equation. In contrast, when an eigen-energy exceeds $t/3$,
the Dirac equation gets poorer to describe such an eigenstate in the
square GQD. From Eq.(\ref{wavevector}) we can see that the interval
between two adjacent wavevectors in the armchair direction is
$\Delta p_n=\pi/L'_x$. The discrete wavevectors in the zigzag
direction should be numerically determined from
Eq.(\ref{transcendental equation}). Therefore, it is not
straightforward to find the interval between the adjacent
wavevectors in this direction. However, for a relatively small $P_n$
we infer from Eq.(\ref{transcendental equation}) that such a
wavevector interval is roughly equal to $\Delta q=\pi/L'_y$. Thus,
the two kinds of wavevector interval are inversely proportional to
the sizes in the respective directions. When $\Delta p_n$ and
$\Delta q$ become comparable to the diameter of the circle of the
linear dispersion limit, there is hardly any eigenstates in the
linear dispersion region. Thus, from the relation $\Delta p_n=\Delta
q=4/(3\sqrt{3}a)$ we can find the minimal size of the GQD for the
complete invalidity of the Dirac equation description. By a simple
evaluation, we find that such a minimal size is $N_{min}=7$ and
$M_{min}=8$. This means that the GQD with size of N7M8 is the
smallest one to which the Dirac equation is applicable.

\begin{figure}
\begin{center}
\scalebox{0.5}{\includegraphics{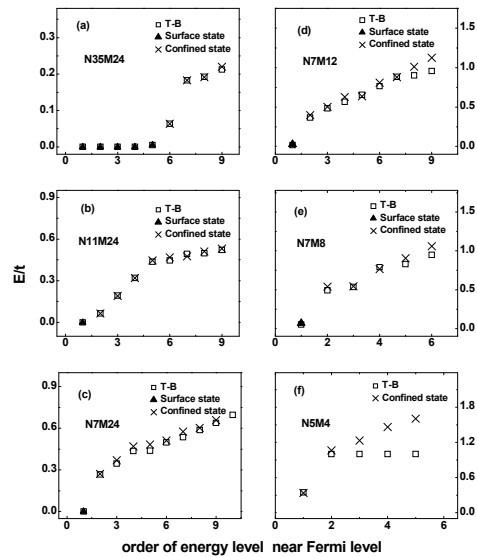}} \caption{A comparison of
the low-lying energy levels calculated from the Dirac equation
(triangle,cross) and tight-binding model (square) for GQDs with
different sizes. (a) N35M24; (b) N11M24; (c) N7M24;(d) N7M12; (e)
N7M8 and (f) N5M4. \label{2}}
\end{center}
\end{figure}

 \par
To check the validity of the Dirac equation solution about the
electronic eigenstate of the GQD in some details, we compare the
low-lying energy levels calculated by solving
Eqs.(\ref{energy2}-\ref{transcendental equation}) to the ones
obtained from the tight-binding model. The tight-binding Hamiltonian
of the square GQD takes a form as $H_{tb}=-t\sum_{\langle
i,j\rangle}\sum_{\sigma}(c^\dag_{i\sss{A}\sigma}c_{j\sss{B}\sigma}+h.c)$
where $c^\dag_{i\mu\sigma}$ is the electron creation operator
associated with a local atomic state at lattice point $i$; And
$\langle i,j\rangle$ denotes any pair of the nearest neighboring
carbon atoms. $\sigma=\uparrow(\downarrow)$ corresponds to the up
and down spins. In the basis set consisting of the local atomic
orbits, the tight-binding Hamiltonian changes into a matrix. By
diagonalizing this Hamiltonian matrix we can obtain the electronic
eigen-energy spectrum and the eigen wavefunctions. Noting that
electronic eigen-states are spin-degenerate, though we write the
spin index explicitly in the above tight-binding Hamiltonian. The
electron spin becomes relevant only in the self-consistent
calculation of the electron energy spectrum in the next section
where the Hubbard interaction is taken into account.
\par
The comparison of the numerical results of the low-lying
eigen-energy spectra obtained by the Dirac equation as well as the
tight-binding model is visualized in Fig.\ref{2}. For a relatively
large GQD, N35M24 as shown in Fig.\ref{2}.(a), the Dirac equation
result agrees with the tight-binding result very well. Then we
reduce the size of the GQD only in x direction. These results are
plotted in Fig.\ref{2}(a)-(c). We can see that the Dirac equation
results get poorer with the decrease of $N$. In Fig.\ref{2}(c),
despite $N=7$, there are a few low-lying eigen-energies obtained by
the Dirac equation solution to be close to the tight-binding result.
This is due to that the relatively large $M$ retains these
eigenstates in the linear dispersion region. When $M$ decreases
further, the results calculated by the two models deviate from each
other notably, as shown in Fig.\ref{2}(c)-(e). The result shown in
Fig.\ref{2}(f) demonstrates that the Dirac equation description
fails to give the correct eigen-energy spectrum for the GQD smaller
than the one of N7M8. On the other hand, the result shown in
Fig.\ref{2}(a) indicates that at least 10 low-lying eigenstates can
be safely described by Dirac equation for a GQD with size of
4nm$\times$5nm. In conclusion, the numerical comparison made in
Fig.\ref{2} supports our simple criterion given above for the
applicable limit of the Dirac equation approach to the GQD.

\begin{figure}
\begin{center}
\scalebox{0.6}{\includegraphics{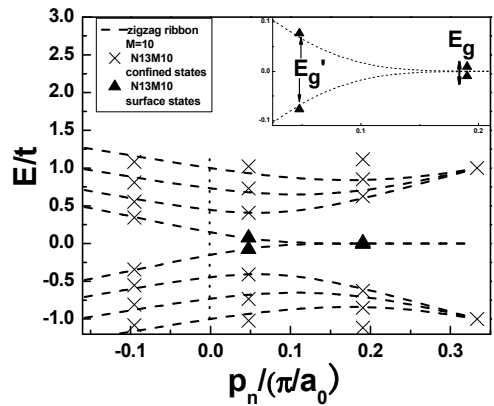}} \caption{ Low-lying
energy levels versus $p_n$ for a GQD of size N13M10 calculated from
Dirac equation (triangle). The dashed lines depict the dispersion
relation of an infinitely long zigzag nanoribbon with width M=10,
which is calculated from the tight-binding model. The inset shows
more clearly the two pairs of the conjugate surface states of the
GQD. And the energy gaps relative to the largest $p_n$ and the
smallest $p_n$ are denoted by $E'_g$ and $E_g$ respectively .
\label{3}}
\end{center}
\end{figure}

\par
In Fig.\ref{3} some low-lying eigen-energies versus $p_n$ are
plotted for a GQD of size N13M10, which is compared with the
dispersion relation of a zigzag ribbon with width M=10. Although
these energy levels of the GQD are discrete, from this figure we can
readily infer that the energy-wavevector relation of a GQD will
change into the band structure of a zigzag nanoribbon with the
continued increase of the size $N$. Thus, we can say that the
dispersion relation of the zigzag nanoribbon remains in the Dirac
equation description of the square GQD. This is one advantage of the
Dirac equation over the tight-binding model in describing the
electron states of the GQD. In addition, from the Dirac equation
solution, we can easily distinguish the surface states from the
confined states, because the two kinds of state have distinct forms
of wavefunction. This can be viewed as another advantage of the
Dirac equation description. In contrast, it is difficult to identify
a surface state in the tight binding model, in particular, for a GQD
with small size. In fact, the Dirac equation method was previously
used to describe the electron states in other
GQDs\cite{trauzette,efetov,martino}. For example, for a GQD formed
by applying a gate voltage on a graphene nanoribbon\cite{trauzette},
the surface states and the confined states can also be easily
distinguished from each other by using the Dirac equation method.
Finally, from the inset of Fig.\ref{3} we can clearly see that
corresponding to any allowed $p_n$, there are two conjugate surface
states, belonging to the valence and conduction bands respectively.
And between them there is a finite energy gap. Although there are
only two pairs of surface states for the GQD shown in Fig.\ref{3}, a
common feature about the surface state visualized in this figure is
that the pair of surface state with the minimal $p_n$ has just the
maximal energy gap and vice versa.
\par
\section{Magnetic properties of the GQD\label{magnet}}
Magnetic properties of graphene nanostructures with various shapes
have drawn considerable interest. For example, some theoretical
investigations indicate that a zigzag nanoribbon has a
spin-polarized ground state, which is tightly associated with the
surface state at the zigzag edges. This implies that a spontaneous
magnetization may occur in such a nonferromagnetic material.
Motivated by these previous work\cite{Jianga,Fern,Oded,pisani}, we
now study the possible magnetic property of the square GQD. To do
this, we adopt a single band Hubbard model(to incorporate the
Hubbard terms into the tight binding model), and treat it within the
Hartree-Fock approximation. It was previously proved that most
magnetic properties of graphene nanostructure can be captured by
such a simple approach\cite{Fern,fujita,pisani}. The Hartree-Fock
mean field Hamiltonian of the present GQD take a form
\begin{equation}
H=H_{tb} +U \sum_{i\mu}(\langle n_{i\mu \downarrow} \rangle n_{i\mu
\uparrow}+\langle n_{i\mu \uparrow} \rangle n_{i\mu \downarrow}-
\langle n_{i\mu \uparrow} \rangle \langle n_{i\mu \downarrow}
\rangle ).
\end{equation}
where $n_{i\mu\uparrow}(=c^\dag_{i\mu\uparrow}c_{i\mu\uparrow})$ and
$\langle n_{i\mu\uparrow}\rangle$ are the electron number operator
and the average electron occupation at an arbitrary lattice point
respectively. Besides, $U$ is the on-site Hubbard energy. For a
charge neutral GQD, the single-electron picture adopted above tells
us that all eigenstates in the valence band are fully occupied,
whereas all states belonging to the conduction band are empty.
However, the finite Hubbard $U$ distorts such a simple electron
distribution since it resists the double occupancy of a surface
state by two opposite-spin electrons. As a result, spin polarized
electron occupancy on individual lattice points may occur in the
charge neutral GQD.
\par
\begin{figure}
\begin{center}
\scalebox{0.5}{\includegraphics{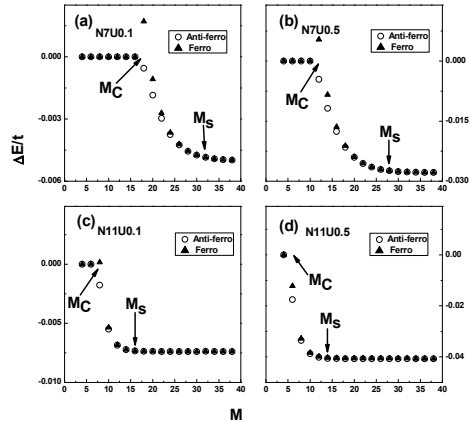}} \caption{Energy
differences versus M between AFM(circle) and PM as well as
FM(triangle) and PM states for charge neutral GQDs with different
sizes and on-site energies. (a)N7U0.1t; (b)N7U0.5t; (c)N11U0.1t; and
(d) N11U0.5t. The critical size $M_c$ and the saturated size $M_s$
are labeled on the curves. \label{4}}
\end{center}
\end{figure}

The eigen solution of the Hartree-Fock Hamiltonian for a charge
neutral GQD can be obtained by iteration method. In analogy with the
previous work, our iterative calculation indicates that the spin
polarization situation of the obtained eigen-state of the charge
neutral GQD depends on the initial spin configuration  to start the
iteration procedure. The different initial states will lead to the
eigen-states with distinct kinds of spin polarization. At first, to
begin with an initial spin configuration of Neel order, we will
arrive at an eigen-state with spin polarized electron occupation on
individual lattice points. In particular, on the lattice points near
the two zigzag edges the spin polarization is very strong. The net
spin distributions at the two zigzag edges show the
anti-ferromagnetic order(AFM) in Fig.\ref{1}. Second, if we start
from an initial state with the uniform spin polarization at all the
lattice points, the self-consistent calculation converges to an
eigen-state with ferromagnetic(FM) ordering. In addition, a
paramagnetic state(PM) can be achieved if the initial spin
configuration is set to be unpolarized at all lattice points. Herein
we adopt the same definition about the magnetic orderings as given
in the previous works\cite{pisani}. The AFM state refers to that the
spin moments of the carbon atoms on one zigzag edge are anti-aligned
to that on the opposite edge, while the FM state means that the spin
moments on both zigzag edges point at the same direction. Moreover,
the PM state can be defined alike, which refers to that the spin up
and down electrons are equally occupied at every lattice point. Next
we focus on the size dependence of the magnetic orderings. To do
this, in Fig.\ref{4} we compare the total energy difference($\Delta
E$) between the FM(AFM) states and the PM state for the charge
neutral GQD, as a function of the width $M$ between the two zigzag
edges, while the width $N$ in the armchair direction takes several
typical values. First of all, we can find that there exists a
critical size $M_c$, which denotes the onset of the PM-FM(AFM)
transition. Namely, only when $M>M_c$ is the magnetic ordering
stable. By comparing the results shown in Fig.\ref{4}(a-d), we find
that $M_c$ depends on the transverse width $N$ and the Hubbard $U$
sensitively. With the increase of $N$ and $U$, the critical size
$M_c$ decreases notably.

\par
There is one point we have to emphasize herein about the magnetic
ordering in the GQD depicted above. The FM or AFM state in the GQD
considered by us and the one in an extended honeycomb lattice have
distinct underlying mechanisms. It is the surface states localized
at the zigzag edges to cause the AFM in the GQD. The surface states
in the GQD tend to dispersionless as the width between the opposite
zigzag edges getting large. Thus, a finite density of states forms
in the vicinity of the Dirac point(energy zero point), which is
responsible for the formation of the magnetic ordering even with a
very small Hubbard U. However, such a surface state is absent in the
extended honeycomb lattice in which there is no zigzag edge. The
linear dispersion of the extended honeycomb lattice leads to the
vanishing density of states at the Dirac point, which then requires
a very large Hubbard U for the PM-AFM transition in such an extended
lattice. In fact, a similar comparison was made in
Ref.\onlinecite{fujita} where the magnetic ordering in a zigzag
ribbon only needs a much smaller(infinitely small, in fact) Hubbard
U than in the extended honeycomb lattice. In analogy with the GQD,
the zigzag ribbon possesses surface states which form a flat band at
the Dirac point, leading to the magnetic ordering even with an
infinitely small U. In short, we think the magnetic ordering formed
by a small U is due to the existence of the surface state in the
GQD. The spin polarized electron occupancy is notable only in the
regions near the zigzag edges. The value of the Hubbard U we used in
our work is too small to show the formation of the AFM state
following the mechanism of the extended lattice(from the result
shown in Ref.\onlinecite{fujita} we can find out that such a
critical U is larger than 2t).

\par
 Another noticeable feature in Fig.\ref{4} is that the energy
difference tends to a saturated value when the width $M$ exceeds a
specific value $M_s$. Hereafter we call $M_s$ the saturated size. It
is quantitatively determined following such a way: when the size of
the GQD increases from $M_s$ to $M_s+2$, the relative increment of
$\Delta E$ should be less than 1\%.  In addition, from Fig.\ref{4}
we can find that the energy of the AFM state is notably lower than
that of the FM state in the region $M_c<M<M_s$. This indicates that
the AFM state in the region between the critical size and the
saturated size is just the ground state of the charge neutral
GQD\cite{Jianga,Lieb}. On the other hand, when $M$ is sufficiently
large, the energy difference between the FM and AFM states becomes
indistinguishable. This can be explained in such a way: The surface
states in the GQD can be viewed as the bonding or anti-bonding
states arising from the interaction between two kinds of surface
states located at the opposite zigzag edges. In fact, each of such
two surface states belongs to a semi-infinite two-dimensional
graphene terminated by a zigzag edge. When $M$ is very large, the
interaction between the two kinds of surface states gets weak. As a
result, the exchange integral between them which determines the
relative orientations of the net spin moments at two zigzag edges
becomes negligibly small. Thus the energy difference between FM and
AMF states tends to zero. Finally, we have to point out that the
existence of the critical size for a hexagonal and square graphene
quantum dots was previously reported\cite{Fern,Jianga,Oded}, based
on the first principle calculation. The result about the square GQD
agrees quantitatively with our present result obtained by the mean
field approximation\cite{Jianga,Oded}. However, a clear explanation
about the critical size is yet lacking.

\par
\begin{figure}
\begin{center}
\scalebox{0.5}{\includegraphics{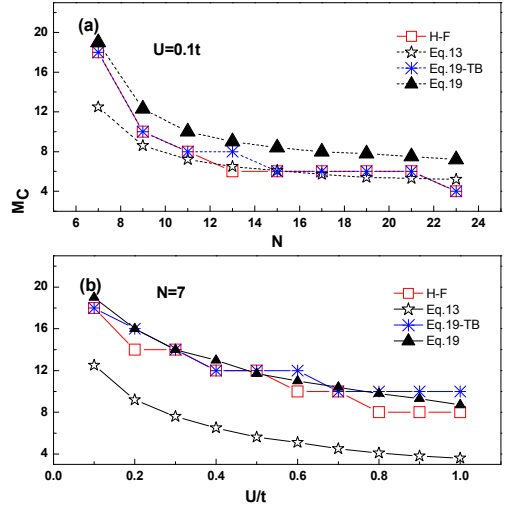}} \caption{ (a) The
critical size $M_c$ versus the transverse size $N$. (b)The critical
size $M_c$ versus $U$. The square symbol denotes the Hartree-Fock
mean field result; The star symbol is the result obtained from the
simple criterion given by Eq.(13). The triangle and cross symbols
for the results obtained from the two-state model. The former is the
result that the parameters $E_g$ and $U_0$ are evaluated from Dirac
equation. And the latter is the result with $E_g$ and $U_0$
calculated from tight binding model. \label{5}}
\end{center}
\end{figure}

Next we try to give a reasonable explanation about the occurrence of
the critical and saturated sizes. At the first step, we discuss the
relation between the magnetic order formation and the electron
occupancy on the surface states. Accordingly, we can work out a
simple criterion which can qualitatively explain the numerical
result about the critical size obtained above by the Hartree-Fock
mean field theory. Then, we establish a two-state model which can
give a quantitative explanation to the numerical result about the
critical and saturated sizes. All these arguments benefit from the
Dirac equation solution about the single-particle surface states. As
analyzed in the preceding section, the number of the surface states
in the range $1/L_y'<p_n\leq \pi/3a_0$ is finite. Corresponding to a
specific $p_n$, there are two conjugate surface states, belonging to
the conduction and valence bands respectively. And the energy gap
between them depends on $p_n$ and the size of the GQD. We consider
specially the two conjugate surface states with the maximal $p_n$.
In comparison with other surface states, this pair of surface states
has the smallest energy gap. The finite Hubbard $U$ alters the
single electron energy spectrum since it affords an on-site Coulomb
repulsive potential. For example, in a presumed paramagnetic state,
the surface state with a specific spin $\sigma$ in the valence band
 will rise by $U\langle
n_{i\bar{\sigma}}\rangle=U/2$ due to the Coulomb repulsion. When
such a shift makes the surface state in the valence band is aligned
with the surface state in the conduction band, the system is likely
to show the spontaneous magnetization. Following such an analysis,
we obtain a simple criterion to determine the critical size.  $M_c$
is the size at which the inequality
  \begin{equation}
  E_g <  U/2\label{critical condition}
  \end{equation}
begins to hold true, where $E_g$ is the energy gap between the pair
of surface states corresponding to the maximal $p_n$. By calculating
$E_g$ we can obtain the critical size $M_c$. Using such a simple
criterion, we estimate the critical size $M_c$, varying with the
 size $N$ as well as the Hubbard $U$. These results are
shown in Fig.\ref{5}. In comparison with the Hartree-Fock mean field
result, we find that the criterion given by Eq.(\ref{critical
condition}) can roughly account for the dependence of the critical
size on $N$ as well as $U$.
\par
Although the above criterion is not too bad, we will show that a
more quantitative explanation about the critical size is available.
Now that the onset of the magnetic ordering is controlled by the
pair of surface states with the minimal energy gap, we establish a
simple two-state model by retaining only this pair of surface states
in the mean field Hamiltonian. Thus, the Hartree-Fock Hamiltonian
for this two-state model takes a form as
\begin{eqnarray}
H^{ts}=\sum_{s\sigma}\varepsilon_{s}c^\dag_{s\sigma}c_{s\sigma} +U
\sum_{i\mu\sigma}\langle \hat{m}_{i\mu \sigma} \rangle \hat{m}_{i\mu
\overline{\sigma}}. \label{Hs}
\end{eqnarray}
here $s=\pm$ denotes the two surface states belong to the conduction
and valence bands respectively. $c_{s\sigma}(c^\dag_{s\sigma})$ is
the electron annihilation(creation) operator of the surface state of
quantum index $s\sigma$. The electron number operator
$\hat{m}_{i\mu\sigma}$ counts only the contributions of the two
surface states to the electron occupancy on the individual carbon
atoms. According to such a meaning, it is associated with the
electron number operator of the two surface states via a relation
\begin{eqnarray}
\hat{m}_{i\mu \sigma}=\sum_{s}|{\psi}^{s}_{\mu}(i)|^2 \Omega\;
c^\dag_{s\sigma}c_{s\sigma},\label{ntwostate}
\end{eqnarray}
here $\Omega=\sqrt{3}a_0^2/2$ is the area of unit cell, and
$|\bm{\psi}^{s}_{\mu}(i)|^2$ is just the probability of the electron
in surface state (s) appearing in the vicinity of the carbon atom
$\mu$ at lattice point $i$, which can be directly calculated from
the analytical wavefunction in Eq.(\ref{total wavefunction}). For
the charge neutral GQD, the two surface states accommodate two
electrons with opposite spins. We thus have
$\sum_{s\sigma}c^\dag_{s\sigma}c_{s\sigma}=2$. In addition, it
should be noted that in such a two-state model we have ignored the
contributions to the average electron occupancy on the individual
lattice points from other occupied single electron eigen-states,
except for the two surface states retained in this model. This is
because that other occupied electron states in the valence band only
provide a spin-unpolarized charge background, which influences the
opposite spin electrons in the two surface states on an equal
footing. Substituting Eq.(\ref{ntwostate}) into Eq.(\ref{Hs}), we
obtain the following diagonal Hamiltonian
\begin{eqnarray}
H^{ts}=\sum_{s,\sigma}(\varepsilon_{s} +\sum_{s'}U_0\langle
c^\dag_{s'\bar{\sigma}}c_{s'\bar{\sigma}} \rangle)
c^\dag_{s\sigma}c_{s\sigma},(s'=\pm s)\label{H2}
\end{eqnarray}
with
\begin{equation}
 U_0=\sum_{\mu}U\cdot\Omega\int dr |{\psi^{s}}_{\mu}(\bm{r})|^2\cdot
 |{\psi}^{s'}_{\mu}(\bm{r})|^2.
\end{equation}
Here $U_0$ can be calculated analytically by the Dirac equation
wavefunction given in Eq.(\ref{total wavefunction}), or numerically
by tight-binding model instead. By a simple derivation from the
Dirac equation wavefunction we obtain the analytical form about
$U_0$. It is given by
\begin{eqnarray}
U_0=\frac{3\sqrt{3}U(\frac{\sinh{4qL_y}}{16q}-\frac{\sinh{2qL_y}}{2q}+\frac{3L_y}{4})}{L_x(\frac{\sinh{2qL_y}}{q}-2L_y)^2}
\end{eqnarray}
\par
For the PM state, two electrons occupy the valence band surface
state($\varepsilon_{-}$) with opposite spins. From the diagonal
Hamiltonian given above, we can immediately obtain that the energy
of the two electrons is equal to $E_{pm}=2\varepsilon_{-}+2U_0$. On
the other hand, for the possible magnetic ordering state, the two
electrons occupy the two distinct suface states with the same spin.
The correspond energy is then
$E_{mo}=\varepsilon_{-}+\varepsilon_{+}$. The critical size for the
magnetic ordering to become stable corresponds to $E_{mo}\leq
E_{pm}$, Namely,
  \begin{eqnarray}
E_g\leq 2U_0. \label{twostate}
\end{eqnarray}
By means of this criterion we can determine the critical size $M_c$.
It should be noted that although the two-state model is established
according to the Dirac equation description of the surface states,
the two parameters $E_g$ and $U_0$ can also be calculated from the
tight-binding model. Therefore, the two-state model is expected to
be still valid even when the conjugate surface states are beyond the
linear dispersion region. The critical size obtained from this
two-state model is shown in Fig.\ref{5}. In comparison with the mean
field result, we find that the two-state model can give a
quantitative explanation about the critical size, as a function of
$U$ and $N$. Besides, in Fig.\ref{5}(a) we also find that the
two-state model with the parameters evaluated from the Dirac
equation can no longer predict the critical size satisfactorily with
the increase of $N$. This can be readily understood. In the
two-state model, we consider the pair of surface states with the
maximal $p_n$, which gets away from the center of the valley with
the increase of $N$. As a result, the Dirac equation becomes poorer
to give a quantitative description about the electron probability
amplitude. Instead of the Dirac equation solution, if we evaluate
the parameters $E_g$ and $U_0$ from the tight binding model, as
shown in Fig.\ref{5} the two-state model always gives a satisfactory
result, which demonstrates that the two-state model has captured the
main mechanism dominating the spin polarization in the GQD. Finally,
we would like to point out that our numerical result about the
critical size shown in Fig.\ref{5} coincides with those obtained by
the first principle calculation or the mean field method in the
previous work\cite{Jianga,Oded,Hideki}. In particular, our
calculation indicates that when $N<7$ no surface state exists, hence
no magnetic ordering occurs. And when $N=7$ the critical size is
$M_c=8$. These quantitative results were also produced in the
relevant work obtained by the first principle
calculation\cite{notes}.  According to solution about the surface
state given in the previous section, we can predict that the
critical size $M_c$ will tend to zero, as the size N becomes
sufficiently large. This is because that there must be a
dispersionless surface state pair when N becomes sufficiently large,
regardless of the size M. Thus, The GQD becomes an infinitely long
zigzag ribbon which always possesses the dispersionless surface
states. As a result, the spontaneous magnetization can occur at an
arbitrarily small M.

\par

\begin{figure}
\begin{center}
\scalebox{0.6}{\includegraphics{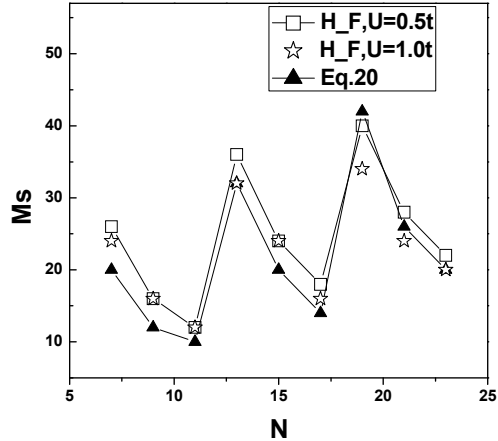}} \caption{ The saturated
size $M_s$ versus the transverse size $N$. The Hartree-Fock mean
field results for U=0.5t(square) and U=t(star) as well as  the
estimated result by using Eq.20 (triangle) are compared. These
results coincide with each other very well. \label{6}}
\end{center}
\end{figure}

Now we turn to discuss the occurrance of the saturated size $M_s$,
based on the Dirac equation description of the surface states. The
mean field result of $M_s$ as a function of $N$ is shown in
Fig.\ref{6}. It depends on $N$ non-monotonously, in contrast to its
insensitive dependence on the Hubbard $U$. For example, when $N$
increases from 7 to 11, $M_s$ decreases notably, followed by an
abrupt rise at $N=13$. Such a situation recurs when $N$ increases
further. According to our solution about the surface state, when $N$
increases within a small range, such as from $N=7$ to 11, the number
of surface states does not change. The pair of surface states with
the minimal $p_n$ has the maximal energy gap(we denoted it as $E'_g$
in the inset of Fig.\ref{3}). According to our two-state model, such
a gap goes against the formation of spin polarization in the two
conjugate surface states. And the energy difference between the spin
polarized and unpolarized electron occupancies on the two surface
states is equal to $E'_g-2U_0$. From
Eqs.(\ref{energy2}-\ref{transcendental equation}) we infer that with
the increase of $M$, $q$ goes close to $p_n$. As a result, $E'_g$
tends to vanish. Meanwhile, as q goes close to $p_n$,
 $U_0$ tends towards a constant, because the wavefunction (Eq.\ref{wavefunction}) of
 surface states near zigzag edges tends
towards a constant value. When $E'_g$ becomes sufficiently small,
such a pair of surface states, hence all pairs of surface states,
have stable contribution to the spin polarization. Thus, the total
energy difference shown in Fig.\ref{4} tends to a saturated value.
In comparison to the case of $N=7$, the GQD with $N=11$ has a
smaller $E'_g$ corresponding to the same $M$. Thus the GQD with
$N=11$

corresponds to a smaller $M_s$. However, when $N$ increases further,
a new surface state comes into being, with an larger $E'_g$ than the
case of $N=7$. Thus a larger $M_s$ is needed. This justifies the
abrupt rise of $M_s$ at $N=13$ as shown in Fig.\ref{6}. In such a
spirit, we can establish a simple criterion by which we can estimate
the value of $M_s$. This is
      \begin{equation}
      p_n-q\leq\eta.\label{saturated condition}
     \end{equation}
where $p_n$ is the smallest wavevector in the surface state allowed
region. $\eta$ is an appropriate small quantity to characterize the
extent that q approaches to $p_n$. The energy gap $E'_g$ is then
sufficiently small and $U_0$ becomes a constant. For the numerical
calculation, we choose $\eta=10^{-4}/a_0$. From Fig.\ref{6} we can
find that such a simple rule creates a saturated size $M_s$ which
agrees with the Hartree-Fock mean field result very well. Finally,
we have to point out that the result in Fig.\ref{6} only shows the
saturated size $M_s$ in a very finite range of the size N, because
that the self-consistent calculation becomes rather time-consuming
as N gets larger. But we can predict that as N increases
furthermore, the oscillation of the $M_s$ will become weak since the
discrete wavevector $P_n$ tends to a continuous quantity. Finally
the saturated size $M_s$ tends to that of the infinitely long zigzag
ribbon.

\section{conclusion and remarks\label{summary}}
The electronic eigenstates of a square GQD terminated by both zigzag
and armchair edges have been studied analytically in the theoretical
framework of Dirac equation. By comparing with the result of tight
binding model, we find that the Dirac equation can well describe the
electron eigen-states even if the size of a GQD reduces to a few
nanometers. Moreover, the Dirac equation method has advantages over
the tight binding model in two aspects. At first, the Dirac equation
solution about the electron eigen-states can tell us not only the
energy levels but also the dispersion relation. Then, from the Dirac
equation solution, we can readily determine the number of the
surface states. In addition, by using the Hartree-Fock mean field
theory, we have also investigated the magnetic properties of the
square GQD. We find that stable magnetic ordering states are allowed
for a charge neutral GQD with an appropriate size. The magnetic
ordering depends on the width between two zigzag edges sensitively.
Only when the width is larger than a critical size is the magnetic
ordering stable. On the other hand, when this width becomes
sufficiently large, the magnetic ordering ground state energy tends
towards a saturated value. We find that the critical size is
dominated by the pair of surface states with the minimal energy gap,
while the saturated size is determined by the pair of surface states
with the maximal energy gap. Based on the Dirac equation
description, we establish a simple model in which only the two
dominated surface states are incorporated. Consequently, this
two-state model can quantitatively explain the size dependence of
the magnetic ordering of the square GQD. Thus, by virtue of such a
toy model, we can estimate rapidly the characteristic sizes for the
formation of magnetic ordering in the GQD.

\acknowledgements This work is supported by the National Natural
Science Foundation of China under Grants No. 10774055 and No.
20771101, also supported by a grant from Hundreds Youth Talents
Program of CAS (Li GS)

\clearpage
%\section{\protect\bigskip\ {\protect\large FIGURES}}

%\end{section}
\end{document}